\documentclass[%
 reprint,
superscriptaddress,
 amsmath,amssymb,
 aps,
 pra,
floatfix,
showkeys,
]{revtex4-2}

\usepackage{graphicx}%
\usepackage{xcolor}
\definecolor{cmblue}{rgb}{0.12156862745098039, 0.4666666666666667, 0.7058823529411765}
\definecolor{myblue}{rgb}{0.2,0.2,0.8}
\usepackage[
colorlinks=true,
citecolor=myblue,
linkcolor=cmblue,
urlcolor=myblue,
pdftitle={From QCCD Trapped-Ion Hardware to Quantum Compilers:  Capabilities, Architectures, and Compilation},
pdfauthor={Ludwig Schmid, Jurek Eisinger, Daniel Schoenberger, Christian D. Marciniak, Janine Hilder, Ulrich Poschinger, Robert Wille, Ferdinand Schmidt-Kaler},
pdfkeywords={trapped ions; quantum charge-coupled device; quantum compilation; ion shuttling; hardware/software co-design; quantum error correction}]{hyperref}
\usepackage[mathlines]{lineno}%
\usepackage{booktabs}
\usepackage{array}
\usepackage{makecell}
\usepackage{threeparttable}
\usepackage[defaultlines=3,all]{nowidow}
\usepackage{braket}
\usepackage{enumitem}

\usepackage{framed}
\usepackage[framed]{ntheorem}
\makeatletter
\newtheoremstyle{summarystyle}
{\item[\theorem@headerfont\hskip\labelsep ##1\theorem@separator]\normalfont}%
{\item[\theorem@headerfont\hskip \labelsep]{\theorem@headerfont ##3}\theorem@separator\newline\normalfont}
\makeatother

\theoremstyle{summarystyle}
\theoreminframepreskip{2pt}   %
\theoreminframepostskip{2pt}  %
\newframedtheorem{summary}{Summary}

\setlist[enumerate]{leftmargin=1.5em, itemsep=2pt, topsep=4pt, parsep=0pt, partopsep=0pt}
\setlist[itemize]{leftmargin=1.5em, itemsep=2pt, topsep=4pt, parsep=0pt, partopsep=0pt}
\setlist[description]{leftmargin=1.5em, itemsep=2pt, topsep=4pt, parsep=0pt, partopsep=0pt}

\AtBeginDocument{\providecommand{\selectlanguage}[1]{}\renewcommand{\selectlanguage}[1]{}}
\usepackage{cleveref}

\begin{document}

\title{From QCCD Trapped-Ion Hardware to Quantum Compilers: \\ Capabilities, Architectures, and Compilation}

\author{Ludwig Schmid}%
\affiliation{Chair for Design Automation, Technische Universität München, 80333 München, Germany}
\author{Jurek Eisinger}%
\affiliation{QUANTUM, Institut für Physik, Johannes Gutenberg Universität, 55128 Mainz, Germany}
\author{Daniel Schoenberger}%
\affiliation{Chair for Design Automation, Technische Universität München, 80333 München, Germany}
\author{Christian D. Marciniak}%
\affiliation{neQxt GmbH, 80686 München, Germany}
\author{Janine Hilder}%
\affiliation{QUANTUM, Institut für Physik, Johannes Gutenberg Universität, 55128 Mainz, Germany}
\affiliation{neQxt GmbH, 63906 Erlenbach am Main, Germany}
\author{Ulrich Poschinger}%
\affiliation{QUANTUM, Institut für Physik, Johannes Gutenberg Universität, 55128 Mainz, Germany}
\affiliation{neQxt GmbH, 63906 Erlenbach am Main, Germany}
\author{Robert Wille}%
\affiliation{Chair for Design Automation, Technische Universität München, 80333 München, Germany}
\affiliation{MQSC, 85748 Garching, Germany}
\author{Ferdinand Schmidt-Kaler}%
\affiliation{QUANTUM, Institut für Physik, Johannes Gutenberg Universität, 55128 Mainz, Germany}
\affiliation{neQxt GmbH, 63906 Erlenbach am Main, Germany}

\begin{abstract}
{Shuttling-based trapped-ion quantum computers (QCCD) rank among the most mature hardware platforms. Existing and envisaged devices exhibit architectural differences in terms of register layout and connectivity. Platform-specific compilers are required to translate quantum circuits into resource-optimized hardware-level command chains. Hardware platforms and compiler software stacks are mostly developed independently of one another by different research communities. This work aims to bridge these communities by abstracting QCCD architectures and native operations, enabling computer scientists to address compilation problems using generic methods without requiring a deep background in the physical operating principles of the platform. The resulting design space is organized into a small set of architectural archetypes, each posing its own compilation problem. We then select and apply compilation methods to one- and two-dimensional QCCD architectures, finding that the routing overhead scales approximately cubically with the number of qubits for the one-dimensional architecture and approximately quadratically for the two-dimensional one. Extending this to compilation with a quantum error-correction layer reveals additional differences and similarities between the physical and logical levels. We conclude that co-design is a requirement rather than an option, and this work provides the vocabulary, archetypes, and compilation abstraction it needs.
}
  
\end{abstract}


\maketitle

\section{Introduction}

Trapped-ion quantum computers have progressed from experimental proof-of-principle demonstrators to programmable processors with dozens of qubits and high-fidelity quantum logic operations~\cite{bruzewiczTrappedIonQuantumComputing2019, mosesRaceTrackTrappedIonQuantum2023, ransford98qubitTrappedionQuantum2026}. Scaling these systems further to application-grade platforms remains an engineering challenge. Most proposals are based on the \emph{quantum charge-coupled device} (QCCD)  architecture, where ions are stored in multiple zones of a segmented trap and dynamically rearranged via \emph{shuttling operations}~\cite{kielpinskiArchitectureLargescaleIontrap2002}. Within this paradigm, concrete designs exhibit a wide range of variation: linear multi-zone traps~\cite{mosesRaceTrackTrappedIonQuantum2023}, two-dimensional layouts with junctions~\cite{blakestadHighFidelityTransportTrappedIon2009}, and large-scale planar arrays~\cite{lekitschBlueprintMicrowaveTrapped2017, malinowskiHowWire1000Qubit2023,valentini2025}. Long-term scalability will most likely require connecting separate processor modules via photonic or material interconnects~\cite{kimModularUniversalScalable2011, monroeLargescaleModularQuantumcomputer2014}.
But none of these designs is final or generally superior: how many processing zones a device should have, where to include junctions, and how much addressing to provide within a zone remain open questions.

On the other hand, as hardware capabilities keep on increasing, the performance of a platform is determined to an equal extent by its control software stack. Any quantum protocol to be executed must first be \emph{compiled}. The compiler translates abstract circuit representations into sequences of native gates~\cite{maslovBasicCircuitCompilation2017}, maps algorithmic qubits onto physical qubits (ions), and generates optimized shuttling schedules~\cite{kreppelQuantumCircuitCompiler2023, durandauAutomatedGenerationShuttling2023, schoenbergerShuttlingScalableTrappedIon2024, webberEfficientQubitRouting2020}, all while respecting device and control constraints and tracking the error budget~\cite{wuMUSSTIMultilevelShuttle2025, sakiMuzzleShuttleEfficient2022}. Its output is lowered to hardware-level instructions representing radiation pulses and voltage waveforms~\cite{kangBatchOptimizationFrequencyModulated2021, sterkClosedloopOptimizationFast2022}.
The optimization of each of these layers presents a challenging problem, and integrating them into one comprehensive toolchain adds a software-engineering challenge on top: currently the field is moving from research-grade solutions toward complete compilation stacks~\cite{schmaleBackendCompilerPhases2022, willeMQTHandbookSummary2024}, increasingly built on modern compiler infrastructure such as Multi-Level Intermediate Representation (MLIR)~\cite{lattnerMLIRScalingCompiler2021, mccaskeyMLIRDialectQuantum2021,hopfIntegratingQuantumSoftware2026}.
This effort pays off directly: powerful compilation routines yield more compact operation sequences, increasing the range of applications that can be targeted with an existing platform~\cite{muraliNoiseAdaptiveCompilerMappings2019, ferrariNoiseadaptiveQuantumCompilation2022}.

The expertise needed to develop such compilers is abundant in computer science, where compiler development is a well-established discipline that draws on decades of experience in the areas of intermediate representations, optimization passes, and toolchain development~\cite{lattnerMLIRScalingCompiler2021}.
Many of the ideas and techniques developed there carry over to compilation tasks for quantum information processing platforms~\cite{mccaskeyMLIRDialectQuantum2021}. A common roadblock for successful interdisciplinary co-design of hardware platform and compiler stacks is an insufficient understanding of the underlying physical operating principles~\cite{bruzewiczTrappedIonQuantumComputing2019, haffnerQuantumComputingTrapped2008}, e.g.\ the available native operations for a given platform and their implementations, and the physical background of the resource budgets the compilation stack is supposed to optimize. Hardware developers face the converse problem: they often commit to architectural decisions before actually building a large-scale device, without consulting quantitative guidance on how these decisions play out once the platform is used for applications.
The purpose of this work is to build a bridge between compiler and hardware developers, to facilitate the development of improved compilation methods which enhance the capabilities of existing platforms and to enable efficient co-design of hardware and software.

This work consists of four steps to establish such a bridge:
\textbf{(1)}~We abstract the elementary operations of QCCD devices, namely gates, state preparation and measurement (SPAM) operations, and shuttling primitives, together with time and fidelity costs, into a compact form that does not require a background in atomic physics (\Cref{sec:platform}).
\textbf{(2)}~We describe how these operations compose into architectures and organize the resulting design space into a small set of archetypes with characteristic strengths and weaknesses (\Cref{sec:architecture}).
\textbf{(3)}~We discuss the compilation steps that these architectures require, for one- and two-dimensional devices, including the additional layer that quantum error correction introduces (\Cref{sec:compilation}).
\textbf{(4)}~We combine both sides in case studies that pair architectures with compilers and extract quantitative guidance for hardware design, illustrating the co-design perspective that motivates this work (\Cref{sec:case_studies}).

\section{Capabilities of the trapped-ion platform}\label{sec:platform}
Trapped ions have historically been among the first platforms to study and control isolated few-body quantum systems~\cite{leibfriedQuantumDynamicsSingle2003}. This has been enabled by the relative ease with which tight electric confinement can be achieved to trap the ions. Moreover, the Coulomb repulsion between the ions can be used to mediate strong long-range interactions, from which one can engineer high-fidelity entangling gates. Since the first proposals and demonstrations~\cite{ciracScalableQuantumComputer2000, schmidt-kalerRealizationCiracZoller2003,leibfriedExperimentalDemonstrationRobust2003}, a considerable variety of approaches to trapped-ion quantum information processing has been developed. We restrict this overview to the most common instances of gate-based quantum information processors, of which an increasing number of platforms from academic~(e.g.\ \cite{eganFaulttolerantControlErrorcorrected2021, zhuGenerationThermofieldDouble2020, pogorelovCompactIonTrapQuantum2021, postlerDemonstrationFaultTolerantSteane2024, mordiniMultizoneTrappedIonQubit2025, clarkEngineeringQuantumScientific2021, revelleRoadrunnerTrapQSCOUT2023, hilderFaultTolerantParityReadout2022}) and commercial~(e.g.\ \cite{mosesRaceTrackTrappedIonQuantum2023, hongEntanglingFourLogical2024, zhaoOrbitaloptimizedPaircorrelatedElectron2023, maReshapingQuantumDevice2025, ollitraultEstimationElectrostaticInteraction2024, ricardoWarehouseOptimizationUsing2024, hilderQuantumInternetNutshell2025}) providers have become available in recent years.

\begin{figure*}[t]
    \centering
    \includegraphics[width=0.95\linewidth]{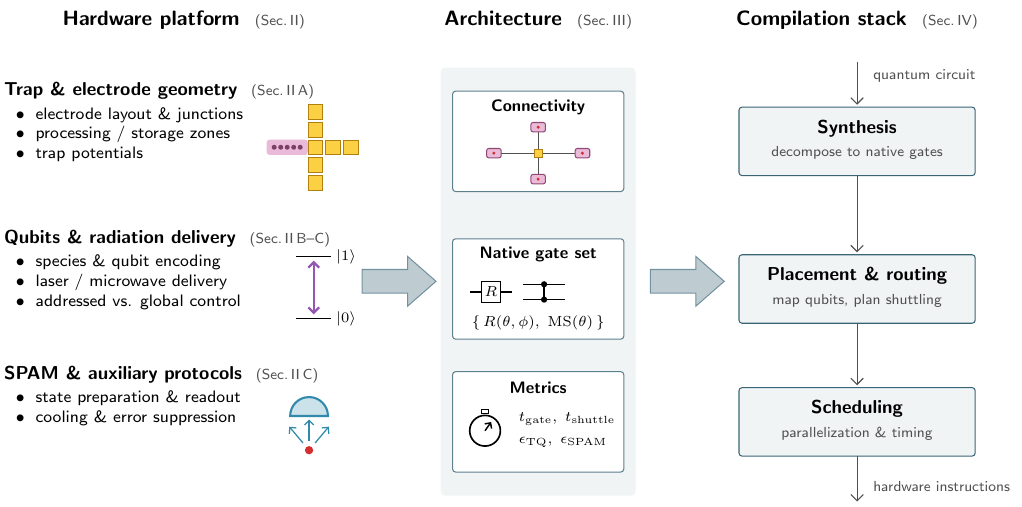}
    \caption{From hardware to architecture: The defining properties of a hardware platform (left), qubit encoding in an ion and delivery of the radiation driving the gates (\Cref{sec:ions_qubits,sec:operations}), and the protocols for SPAM and auxiliary operations (\Cref{sec:operations}), are abstracted into the data sets that define an \emph{architecture} (center): the connectivity (cf.\ the architecture archetypes in \Cref{fig:architectures}), the native gate set, and the operation timings and error budgets.
    The first two follow from one hardware property each, whereas the timings and error budgets are collected from all three, since the gate times are set by the radiation delivery and the shuttling times by the trap geometry. 
    Each data set is then consumed by a different layer of the compilation stack (right, \Cref{sec:compilation}): the native gate set by synthesis, the connectivity by placement and routing, and the timings and error budgets by scheduling.}
    \label{fig:HW2architecture}
\end{figure*}

The sections below describe the hardware in the order of \Cref{fig:HW2architecture}, which also shows how each property is abstracted into architecture data and handed to the compiler layers of \Cref{sec:compilation}.

\subsection{Ion trap types}

To control and manipulate trapped-ion qubits, stable and tight spatial confinement is required.
This is provided by a radio-frequency (rf) trap~\cite{paulElectromagneticTrapsCharged1990}, in which static and oscillating rf electric fields are applied through a suitable electrode structure. The superposition of the rf and static fields produces an effective potential dubbed pseudo-potential, which can exhibit minima in which ions and ion chains can be confined. Terms such as \emph{trap}, \emph{potential well}, and \emph{ion chain} are not always used consistently, which can lead to confusion. To avoid ambiguity, we refer to \Cref{tab:vocabulary}, defining common trapped-ion terms.

Ion traps differ in size, fabrication, topology, and geometry. A common variant is the \emph{blade trap}, which traps a linear ion chain in a single well using a simple, macroscopic electrode geometry~\cite{fsk2003}, a notable example being the Innsbruck Pine trap~\cite{pogorelovCompactIonTrapQuantum2021}.

A \emph{segmented trap} (also: multi-zone or microstructured trap) consists of an array of controllable electrodes and may simultaneously host several wells, which is the key principle of the QCCD architecture~\cite{kielpinskiArchitectureLargescaleIontrap2002}. Each well may hold several ions. Depending on whether gate operations can be performed in a region, wells are referred to as \emph{storage zones} or \emph{processing zones}.

The dimensionality of the electrode geometry must be distinguished from that of the register arrangement. 

In a \emph{linear (1D)} layout, the wells are arranged along a single axis. Qubit connectivity therefore requires rearrangement via logical or physical~\cite{kaufmannFastIonSwapping2017} SWAP operations. A \emph{two-dimensional (2D)} segment layout instead lets ions shuttle through trap junctions, enabling more efficient connectivity. The above distinctions fix how qubits can be brought together to interact, i.e.\ the connectivity and movement model a compiler is based on.

\subsection{Ions and qubit encoding}\label{sec:ions_qubits}
While the trap defines how and where ions can be stored, the qubit itself is encoded in the atomic structure of a specific atomic species. Atomic systems, unlike man-made ones, are provided by nature and are therefore perfectly identical in their fundamental properties. Trapped-ion registers consequently do not suffer from inter-qubit fabrication variances: to a good approximation, all qubits of a given species are interchangeable, and the model handed to a compiler can treat them as such.

However, the atomic properties of the species set hard performance limits, some of which can be mitigated using methods from the atomic-physics toolbox. Quantum information is encoded in two electronic states of an ion, most commonly some isotope of a singly-charged alkaline-earth or alkaline-earth-like atom such as $^9$Be$^+$, $^{40}$Ca$^+$, $^{137}$Ba$^+$, or $^{171}$Yb$^+$~\cite{haffnerQuantumComputingTrapped2008}, whose level structure is sufficiently simple to admit precise radiation-driven operations.
A given atomic species admits different qubit encodings: \emph{ground-state qubits} use Zeeman or hyperfine sublevels of the electronic ground state. \emph{Optical qubits} place one computational basis state in a metastable excited state, while \emph{metastable qubits} encode both computational basis states in metastable electronic levels. Different encodings feature different advantages and drawbacks. Therefore, advanced platforms combine several species~\cite{marinelliRepeatedMultiqubitReadout2019}, or switch the encoding within a single species during operation~\cite{allcockOmgBlueprintTrapped2021}, assigning distinct roles such as data and auxiliary qubits. 

Qubit operations require tight spatial localization of the ions, which in turn requires cooling. Trapped-ion platforms employ laser cooling, causing the ions to form rigid crystalline structures, determined by the balance between the confining potential and the mutual Coulomb repulsion. The most demanding qubit operations require cooling close to the motional ground state: Doppler cooling reaches about a millikelvin, and resolved-sideband cooling tens of microkelvin~\cite{diedrichLaserCoolingZeroPoint1989, monroeResolvedSidebandRamanCooling1995}. Importantly, from an architectural point of view, laser cooling does represent a considerable resource load in state-of-the-art platforms~\cite{Pino2021,hilderFaultTolerantParityReadout2022}. All ions within a crystal are tightly coupled through their shared motion, which mediates the interactions exploited by two-qubit gates, as described in the following section.

\begin{table}[tb]
    \centering
    \caption{Terms for the trap and the ions it holds, in the senses used throughout this work, where parenthesized words may be omitted. Usage varies across the literature.}
    \label{tab:vocabulary}
    \setlength{\tabcolsep}{5pt}
    \renewcommand{\arraystretch}{1.15}
    \footnotesize
    \begin{tabular}{@{}ll@{}}
        \toprule
        Term & Meaning\\
        \midrule
        trap & the electrodes and the fields they produce\\
        (trap) segment & an electrode set defining a section of the trap\\
        zone & one or more segments with an assigned role\\
        \addlinespace
        (potential) well & a minimum of the confining potential\\
        (ion) crystal & the ions held in one well\\
        (ion) chain & an ion crystal in a linear arrangement\\
        \addlinespace
        register & all ions that encode qubits\\
        \bottomrule
    \end{tabular}
\end{table}

\begin{figure*}[t]
    \centering
    \includegraphics[width=\linewidth]{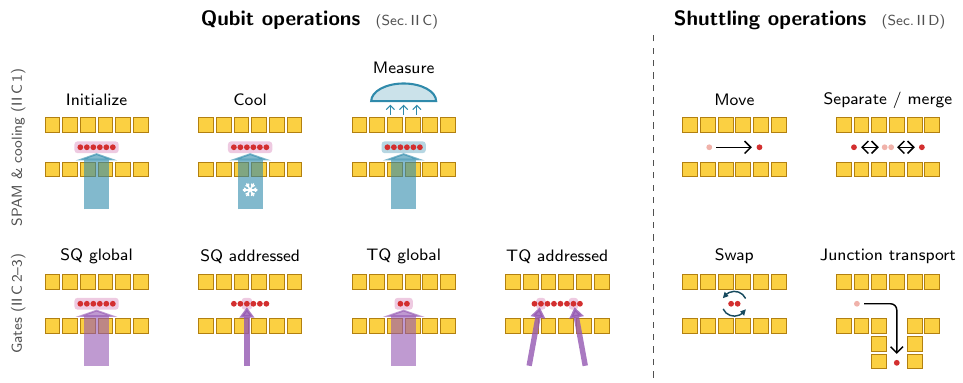}
    \caption{Trapped-ion operations: The qubit operations (left) are grouped by the subsections that describe them. The upper row collects the operations of \Cref{sec:measurement}, where the register is initialized by cooling combined with optical pumping, and read out by fluorescence detection. The lower row shows the single-qubit (SQ, \Cref{sec:1Q_gates}) and two-qubit (TQ, \Cref{sec:2Q_gates}) gates in their global and addressed variants. Purple beams indicate radiation driving the qubit transition, blue beams auxiliary light for pumping, cooling, and detection, with a snowflake singling out the cooling light; strips mark the ions an operation acts on.
    The shuttling operations (right, \Cref{sec:shuttling}) move ions within microstructured traps to establish connectivity; faded dots indicate ion positions before the respective operation.
    Junctions are not restricted to grid crossings, as illustrated by the three-way junction.}
    \label{fig:capabilities}
\end{figure*}

\subsection{Processor operations}\label{sec:operations}
Executing quantum circuits on a trapped-ion platform requires a toolbox of operational primitives. This includes preparing and reading out the register, performing single- and two-qubit quantum logic gates, and shuttling ions to establish the required connectivity. These operations are illustrated in \Cref{fig:capabilities} and discussed in the following.

For gate-based quantum logic, we will focus on the more common case of laser-driven gates, but note that alternative paradigms using microwaves with near-field interactions~\cite{ospelkausMicrowaveQuantumLogic2011, smithSingleQubitGatesErrors2025} or strong magnetic field gradients~\cite{mintertIonTrapQuantumLogic2001, piltzVersatileMicrowavedrivenTrapped2016} also exist. Note that while the choice of the gate-driving radiation heavily impacts the hardware design, compilation software may be designed to be agnostic to such implementation details throughout all hardware-abstracted layers.

Below, each operation is accompanied by a summary box that carries the compiler-facing reading of the physics it describes, under three headings: the \emph{operation} the hardware provides, the \emph{constraint} it imposes on when that operation may be used, and its \emph{cost} in terms of time and incurred error.
Readers interested in compiler constraints rather than the underlying physics can follow the boxes alone; they provide the abstraction that \Cref{sec:architecture,sec:compilation} build on.

\subsubsection{State preparation and measurement (SPAM)}\label{sec:measurement}
Throughout operation, the qubits within the register need to be reliably prepared in a clearly defined state and also measured. These operations are called SPAM operations.

\emph{State preparation} in a well-defined initial state, which is typically the computational basis state $\ket{0}$, is commonly implemented via optical pumping, where population in all other electronic states is depleted via suitable laser radiation.

\emph{Measurements} discriminate the computational basis states $\ket{0},\ket{1}$ and are commonly implemented via state-selective laser fluorescence~\cite{leibfriedQuantumDynamicsSingle2003}, which is collected using imaging optics and detected by an imaging device. Measurements are typically destructive. Therefore, reusing qubits after readout requires subsequent preparation operations.  Such \emph{mid-circuit measurements}, followed by gates conditioned on the measurement results, are required for quantum error correction, and have been demonstrated by several hardware providers~\cite{marinelliRepeatedMultiqubitReadout2019, pelofskeHighRoundQAOAMAX2023, postlerDemonstrationFaultTolerantSteane2024, hothemMeasuringErrorRates2025}.
Isolating the measured ion from the rest of the register renders them expensive, and that cost is paid once per syndrome-extraction round for the entire duration of a fault-tolerant computation (\Cref{sec:fault-tolerance-compilation}).

\begin{summary}[State preparation and measurement]
    \textbf{Operation:} state preparation into $\ket{0\dots0}$ via optical pumping, and readout in the computational basis via state-selective fluorescence.\\
    \textbf{Constraint:} readout is destructive and global on a crystal, so qubit reuse requires re-initialization. 
    Mid-circuit measurement requires isolating the measured ion(s).\\
    \textbf{Cost:} slow compared to gate operations.
\end{summary}

\subsubsection{Single-qubit gates}\label{sec:1Q_gates}
Single-qubit (SQ) gates realize local unitary gates of the forms
\begin{align}
    R_{x,y}(\theta,\phi) &= \exp\left[-\tfrac{i}{2}(\cos\phi\,\hat{\sigma}_x+\sin\phi\,\hat{\sigma}_y)\theta\right] \nonumber \\
    R_z(\theta) &= \exp\left[-\tfrac{i}{2}\hat{\sigma}_z\theta\right]
\end{align}
where $\hat{\sigma}_i,\; i=x,y,z$ are the single-qubit Pauli operators. The reason that we distinguish $x,y$-rotations $R_{x,y}(\theta,\phi)$ and $z$-rotations $R_z(\theta)$ is that they are typically realized differently at the physical level. The $R_{x,y}(\theta,\phi)$ gates are implemented via exposure to resonant radiation driving the qubit transition $\ket{0} \leftrightarrow \ket{1}$. Depending on the qubit encoding used, this is realized by direct optical driving of optical transitions, two-photon stimulated Raman transitions~\cite{leibfriedQuantumDynamicsSingle2003}, radio-frequency, or microwave radiation. The gate angle parameters $\theta,\phi$ are controlled via intensity and phase of the radiation and by the pulse durations.

SQ gates can be applied \emph{globally} on an ion chain exposed to the radiation, meaning all qubits within this (sub-)register undergo the same local gate operation, as indicated by a wide beam covering the whole chain in \Cref{fig:capabilities}. Global control requires no per-ion steering and thus carries low, constant overhead, but is restrictive, since every qubit within a chain undergoes the same gate operation. \emph{Addressed} SQ gates affect only selected qubits within the (sub-)register exposed to the radiation and require laser beams to be tightly focused and steered to single ion qubits.  By contrast, architectures based on long-wavelength radiation (microwave or radio-frequency) do not permit tight focusing, which requires a spatial dependence of the qubit transition frequency e.g.\ via strong magnetic field gradients for frequency-domain addressing.

Several ways exist to realize $z$-rotations $R_z(\theta)$: They can be decomposed into a sequence of $x,y$-rotations or driven directly with off-resonant radiation. Some platforms implement \emph{virtual} $z$-rotations via bookkeeping~\cite{mckayEfficientGatesQuantum2017}, where an $R_z(\theta)$ gate is never physically applied, but effectively included by modifying the phase angle $\phi$ of all downstream $x,y$-gates acting on the same qubit. This enables $R_z$ rotations at zero resource cost, but requires an additional compiler pass to track the phase offsets and modify the subsequent gates accordingly.

\begin{summary}[Single-qubit gates] 
    \textbf{Operation:} native set $\left\{R_{x,y}(\theta,\phi),R_z(\theta)\right\}$, generating any single-qubit unitary.\\
    \textbf{Constraint:} applied globally to a whole chain, or addressed onto a subset or a single ion, depending on the hardware setup and qubit encoding.\\
    \textbf{Cost:} generally fast and high-fidelity; $R_z(\theta)$ can be realized virtually, at zero time and error.
\end{summary}

\subsubsection{Two-qubit gates}\label{sec:2Q_gates}
Gate-based quantum logic ultimately requires two-qubit (TQ) gates implementing nonseparable unitary operations on two selected qubits. This in turn requires some controllable physical interaction between the qubits of interest. For most instances of trapped-ion platforms, this interaction is mediated by the shared motion of ions in a single crystal. Therefore, the two participating qubits are typically required to be part of the same Coulomb crystal. The motion-induced gate coupling also entails the requirement of near-perfect quantum control over motional degrees of freedom, which in turn requires preparing \emph{and} maintaining at least some of these close to the quantum mechanical ground state. 

The TQ gates primarily employed are geometric phase gates, which come in two types:
\begin{align}
    R_{ZZ}(\theta) &= \exp\left[-\tfrac{i}{2}\, \theta\, \hat{\sigma}_{z}^{(i)}\otimes \hat{\sigma}_{z}^{(j)}\right] \nonumber \\
    R_{\phi\phi}(\theta) &= \exp\left[-\tfrac{i}{2}\, \theta\, \hat{\sigma}_{\phi}^{(i)}\otimes \hat{\sigma}_{\phi}^{(j)}\right].
\end{align}
The indices $i,j$ identify the qubits, and $\hat{\sigma}_{\phi}=\cos\phi\hat{\sigma}_x+\sin\phi\hat{\sigma}_y$. Both gates are maximally entangling for $\theta=\pi/2$.
$R_{ZZ}$ gates are commonly called $ZZ$ or light-shift gates~\cite{leibfriedExperimentalDemonstrationRobust2003}, while the $R_{\phi\phi}$ gates are termed M\o{}lmer-S\o{}rensen gates~\cite{sorensenEntanglementQuantumComputation2000}. The two gate types have different physical realizations but are locally equivalent (i.e.\ equivalent up to SQ gates). Furthermore, both maximally entangling TQ gates are locally equivalent to CNOT gates. Some combinations of radiation types and qubit encodings lead to viable gate implementations, while some combinations are not possible or at least unfavorable. For example, M\o{}lmer-S\o{}rensen gates can be implemented via direct bichromatic driving of optical qubits or via stimulated Raman transitions on ground state qubits, while light-shift gates are usually employed in conjunction with magnetic-field-sensitive ground-state qubits and stimulated Raman transitions.

TQ gates are commonly carried out either by global illumination of two-ion crystals, or via multi-beam laser addressing. The latter supports arbitrary $2$-out-of-$N$ gates on longer chains and therefore enables full addressability on this (sub-)register~\cite{figgattParallelEntanglingOperations2019, shapiraTheoryRobustMultiqubit2020}.

Illuminating larger registers globally without addressing would drive simultaneous pairwise entangling gates between all ions within the crystal, which is generally not desired. However, parallel execution of TQ gates within a single crystal would be favorable in terms of runtime. The shared motion mediating the physical interaction imposes a limit on parallelization, as the undesired inter-pair coupling needs to be suppressed. Parallel entangling gates therefore require elaborate modulation schemes~\cite{figgattParallelEntanglingOperations2019}, the use of orthogonal motional modes of one chain~\cite{zhuPairwiseParallelEntanglingGates2023}, or splitting the crystal so that the motion is no longer shared. Parallel addressing is limited further by the radiation delivery hardware: addressing systems generally introduce \emph{crosstalk errors}, i.e.\ undesired coupling of radiation to spectator qubits in close proximity to addressed ions. Such errors will generally increase with the number of simultaneously addressed ions, along with other hardware-specific errors.

Beyond parallel pairwise TQ gates, multi-qubit gates natively acting on more than two qubits have also been proposed and demonstrated~\cite{wangMultibitGatesQuantum2001, ariasespinozaHighfidelityMethodSinglestep2021, goelNativeMultiqubitToffoli2021, katzDemonstrationThreeFourbody2023}. Such gates have so far seen little practical use, due to the additional complexity of implementation and control.

\begin{summary}[Two-qubit gates] 
    \textbf{Operation:} one native entangling gate, the light-shift gate ($R_{ZZ}(\theta)$) or the M\o{}lmer-S\o{}rensen gate (e.g.\ $R_{XX}(\theta)$); both are maximally entangling for $\theta=\pi/2$, and then equivalent to the CNOT gate up to single-qubit rotations.\\
    \textbf{Constraint:} the two ions must share a single crystal, and must be isolated from the other ions, which requires either a separate crystal, orthogonal motional modes, or selective addressing.\\
    \textbf{Cost:} the slowest gate and the largest gate error source.
\end{summary}

\subsection{Shuttling}\label{sec:shuttling}
The currently leading approach to scale trapped-ion platforms beyond static ion chains consists of storing ions in multiple, spatially separated confining potential wells within segmented ion traps. Connectivity is enabled via \emph{shuttling operations}, where ions are controllably moved between segments within the trap. These operations are physically realized by applying suitable time-dependent voltage waveforms to the trap electrodes.
This approach was first proposed in~\cite{winelandExperimentalIssuesCoherent1998a, kielpinskiArchitectureLargescaleIontrap2002} and is dubbed the QCCD architecture, in analogy to modern digital imaging devices.

The shuttling operations can be categorized into four primitives, shown on the right side of \Cref{fig:capabilities}.
A \emph{movement} operation carries a potential well, together with the ions it holds, from one trap segment to another~\cite{waltherControllingFastTransport2012, bowlerCoherentDiabaticIon2012}, while \emph{separate} and \emph{merge} operations change the number of wells instead, splitting one crystal into two or fusing two into one~\cite{rusterExperimentalRealizationFast2014}. A \emph{physical SWAP} rotates a crystal in place and reverses the ion ordering within the crystal~\cite{kaufmannFastIonSwapping2017,kaushalShuttlingbasedTrappedionQuantum2020}. This is to be contrasted with a logical SWAP mediated by quantum logic gates, which acts on the qubits themselves and therefore typically incurs additional errors. Finally, a \emph{junction transport} carries an ion around a corner of a two-dimensional segment layout~\cite{blakestadHighFidelityTransportTrappedIon2009}. An overview of the hardware requirements specific to shuttling-based architectures can be found in~\cite{kaushalShuttlingbasedTrappedionQuantum2020}. Current cutting-edge platforms rely on shuttling to establish full connectivity in registers of several tens of qubits~\cite{mosesRaceTrackTrappedIonQuantum2023,ransford98qubitTrappedionQuantum2026}.
Since each shuttling operation takes place on a distinct, localized set of trap electrodes, shuttling operations in separate trap regions can in principle run simultaneously. The achievable degree of shuttling parallelism is thus determined by the trap layout and the number of independent control voltage channels.

Turning a shuttling-based device into a freely programmable quantum processor requires additional layers in the compilation stack, which are described in \Cref{sec:compilation-physical}.

\begin{summary}[Shuttling] 
    \textbf{Operation:} voltage-controlled moves of ions: transport, separation and merging of crystals, physical SWAPs, and junction transport, together providing effective full connectivity in QCCD devices.\\
    \textbf{Constraint:} moves in separate trap regions can run in parallel, but operations sharing electrodes or the same junction must be serialized, bounded by the number of independent voltage-control channels.\\
    \textbf{Cost:} adds idle time that increases dephasing, plus excess motion that degrades subsequent gates unless the ions are recooled.
\end{summary}

\subsection{Error models}\label{sec:errors}
Error models, expressing which types of errors are induced by elementary operations, are rather platform-specific.
These models serve several purposes: for a given platform, they can be used to assess the capability limitations, compute resource requirements for a given computational task, and identify or design suitable error suppression techniques. Moreover, the error model can serve as input for compilation layers, such that \emph{noise-aware compilation} strategies~\cite{muraliNoiseAdaptiveCompilerMappings2019, ferrariNoiseadaptiveQuantumCompilation2022} can be used to find operation sequences yielding optimized success probabilities.

Error models come at various levels of detail, ranging from the mere specification of a success probability per gate or gate layer down to precise microscopic noise models. For trapped-ion platforms, some specific details should be kept in mind when setting up a suitable error model:
\begin{itemize}
    \item 
    Error parameters due to miscalibration or parameter drift are attached to locations (gate zones and/or addressing channels) rather than qubits.
    \item
    Crosstalk errors describe the undesired impact of gate operations on spectator qubits. Both spatial and spectral proximity may contribute.
    \item 
    Ground-state qubit encodings feature practically unlimited population lifetimes ($T_1$), and no depolarization occurs during idle times. 
    \item
    Depending on the qubit species and encoding, the coherence lifetimes ($T_2$) can vary over a wide range. 
    \item
    Shuttling operations generally inflict no direct errors on the qubits, but contribute indirectly to the effective error rates: the induced timing overhead increases decoherence, and residual motion delocalizes the ions over the driving fields of subsequent gate operations, degrading the gate fidelities.
    In the simplified error model below these contributions are summarized into an effective fidelity, formally attributed to the shuttling operation itself. 
\end{itemize}
These points illustrate some trapped-ion-specific options and challenges for noise-aware compilation stacks: given a sufficiently detailed error model, the compiler may switch between different qubit encodings throughout execution of a circuit, in order to harness advantages and avoid disadvantages of specific qubit types. Furthermore, effective error models for shuttling operations may assist the routing and placement compilation layers in finding near-optimal, noise-informed solutions. 

\begin{summary}[Error model]
    As a first-order proxy we consider an \emph{approximate success probability} $\mathcal{P}$, given by the product of the effective fidelities $\mathcal{F}_{O_i}$ of all operations $O_i$ in a computation $U$, shuttling included, combined with an idle term for the decoherence during idling:
    \begin{equation*}
        \mathcal{P}(U) = \exp \left(- \frac{t_\mathrm{idle}}{T_\mathrm{eff}} \right) \prod_{O_i} \mathcal{F}_{O_i}
    \end{equation*}
    Here $t_\mathrm{idle}$ is the \emph{idle time} summed over all qubits to which no operation is applied, and $T_\mathrm{eff}$ is an effective coherence time that has to be fitted to experimental data rather than derived from first principles.
\end{summary}

\section{Architectures}\label{sec:architecture}
The operational capabilities and constraints discussed in the preceding sections are the building blocks of a trapped-ion \emph{architecture}. Seen from the perspective of \emph{hardware engineers}, an architecture consists of the specification of the qubits and the hardware components. From the perspective of \emph{software engineers}, however, an architecture may be regarded as a complete set of data structures representing the operations, capabilities, and constraints of the hardware platform, which is necessary and sufficient for designing and operating the compiler stack~\cite{schmidComputationalCapabilitiesCompiler2024}. The interplay between these perspectives is depicted in \Cref{fig:HW2architecture}.
In this view, an architecture collects the operation abstractions of \Cref{sec:platform} and fixes how the qubits they act on are arranged and connected, which is what turns a list of capabilities into a machine a compiler can target.

\begin{figure*}[t]
    \centering
    \includegraphics[width=1\linewidth]{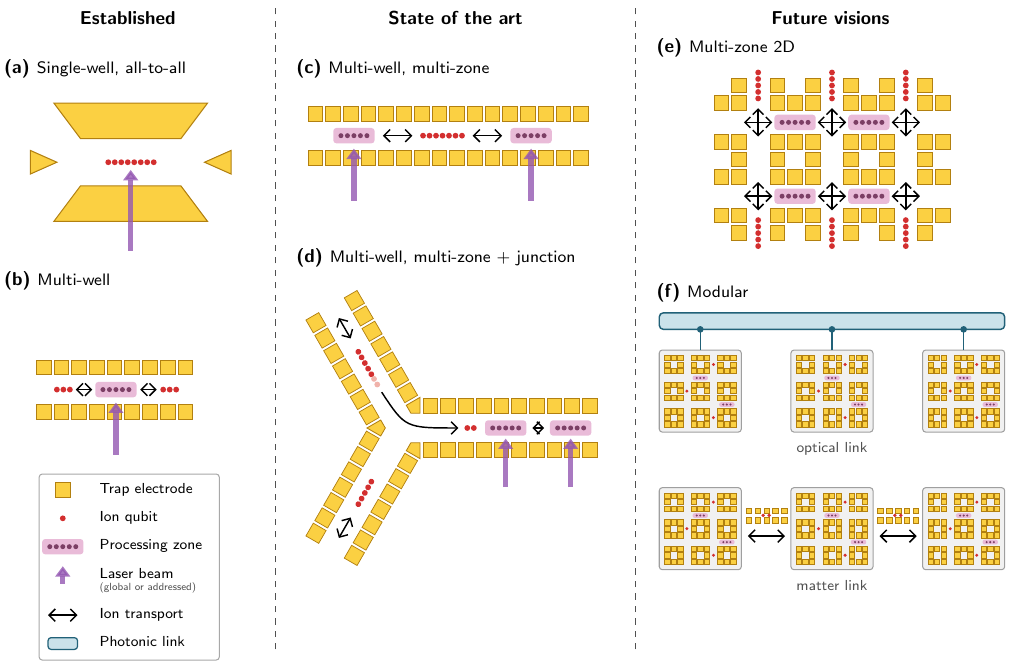}
    \caption{
    Trapped-ion architecture archetypes, grouped by how far each has been realized and named as in \Cref{tab:archetypes}.
    Yellow boxes indicate trap electrodes; multiple of them, together with the rf, generate a confining micro-potential in which trapped-ion qubits (red dots) are stored.
    Processing zones, laser beams, ion transport, and photonic links are marked according to the legend.
    See \Cref{tab:archetypes} for representative realizations and the compilers written for each archetype.}
    \label{fig:architectures}
\end{figure*}

\begin{table*}[t]
    \centering
    \begin{threeparttable}
    \caption{Architecture archetypes of \Cref{fig:architectures}, grouped as in that figure and indexed against the main design axes, the key task each places on the compiler stack, and representative trapped-ion compilers targeting each.}
    \label{tab:archetypes}
    \renewcommand\cellalign{tl}
    \renewcommand\theadalign{tl}
    \setlength{\tabcolsep}{8pt}
    \renewcommand{\arraystretch}{1.15}
    \small
    \begin{tabular}{@{}lllll@{}}
        \toprule
        \thead{Archetype\\(\Cref{fig:architectures})} & \thead{Layout /\\zones} & \thead{Connectivity\\mechanism} & \thead{Key\\compiler task} & \thead{Existing representative\\compilers}\\
        \midrule
        \multicolumn{5}{@{}l}{\textit{Established}}\\[2pt]
        \makecell{Single-well,\\all-to-all (a)} & \makecell{1D chain,\\single trap} & \makecell{laser addressing\\(all-to-all)} & \makecell{gate decomposition,\\crosstalk suppression} & \makecell{Nebendahl~\cite{Nebendahl2009}, Martinez~\cite{Martinez2016},\\ Maslov~\cite{maslovBasicCircuitCompilation2017}}\\
        \addlinespace
        Multi-well (b) & \makecell{1D,\\single zone} & \makecell{1D shuttling,\\global gates} & shuttle scheduling & \makecell{Kreppel~\cite{kreppelQuantumCircuitCompiler2023}, Durandau~\cite{durandauAutomatedGenerationShuttling2023},\\Durandau~\cite{durandau2026heuristicsshuttlingsequenceoptimization}}\\
        \midrule
        \multicolumn{5}{@{}l}{\textit{State of the art}}\\[2pt]
        \makecell{Multi-well,\\multi-zone (c, d)} & \makecell{1D multi-zone,\\(d) adds\\one junction} & \makecell{1D shuttling,\\junction routing} & \makecell{qubit placement,\\routing} & \makecell{Muzzle~\cite{sakiMuzzleShuttleEfficient2022}, Schmale~\cite{schmaleBackendCompilerPhases2022},\\LLM (fine-tuned)~\cite{kreppelShuttlingCompilerTrappedIon2025}, RL~\cite{schierReinforcementLearningIon2026},\\DeCross~\cite{decrossQubitReuseCompilationMidCircuit2023}, Durandau~\cite{durandau2026shuttlingbidimensionalsegmentediontrap},\\Pos.-graph$^\dagger$~\cite{bachEfficientCompilationShuttling2026}, BOSS$^\dagger$~\cite{wuBOSSBlockingAlgorithm2025},\\Ovide$^\dagger$~\cite{ovideAdaptiveParallelismAwareQubit2026}}\\
        \midrule
        \multicolumn{5}{@{}l}{\textit{Future visions}}\\[2pt]
        Multi-zone 2D (e) & \makecell{2D grid,\\many zones} & \makecell{2D shuttling,\\multiplexed control} & \makecell{circuit partitioning,\\parallel scheduling} & \makecell{Ion-shuttler~\cite{schoenbergerShuttlingScalableTrappedIon2024, schoenbergerOrchestratingMultiZoneShuttling2025},\\SAT (exact)~\cite{schoenbergerUsingBooleanSatisfiability2024}, TrapSIMD~\cite{ruanTrapSIMDSIMDAwareCompiler2025a},\\LLM (frontier)~\cite{kreppel2026}, Webber~\cite{webberEfficientQubitRouting2020},\\Pos.-graph$^\dagger$~\cite{bachEfficientCompilationShuttling2026}, BOSS$^\dagger$~\cite{wuBOSSBlockingAlgorithm2025},\\Ovide$^\dagger$~\cite{ovideAdaptiveParallelismAwareQubit2026},\\TISCC$^\ddagger$~\cite{leblondTISCCSurfaceCode2023}, Jones$^\ddagger$~\cite{jonesArchitectingScalableTrapped2026},\\Moveless$^\ddagger$~\cite{khanMovelessMinimizingQEC2025}, Watkins$^\ddagger$~\cite{watkinsHighPerformanceCompiler2024}}\\
        \addlinespace
        Modular (f) & \makecell{modular\\nodes} & \makecell{optical or\\matter links} & \makecell{hierarchical\\routing} & \makecell{Jeng~\cite{jengModularCompilationQuantum2025},\\MUSS-TI~\cite{wuMUSSTIMultilevelShuttle2025}}\\
        \bottomrule
    \end{tabular}
    \begin{tablenotes}
    \footnotesize
    \item[$\dagger$] Largely architecture-agnostic; applies across several rows.
    \item[$\ddagger$] Surface-code compiler, operating at the logical layer.
    \end{tablenotes}
    \end{threeparttable}
\end{table*}

The fundamental challenge in designing architectures lies in enabling scalability at a realistic level of hardware complexity, while retaining sufficiently high operational fidelities. Some archetypes of trapped-ion architectures at different levels of capabilities and implementation maturity are sketched in \Cref{fig:architectures}. At \emph{low- to intermediate scalability}, shuttling operations are required to establish effective connectivity among larger structures. For future \emph{large-scale} platforms, additional building blocks and operations may be required, such as exchanging quantum information between different trap modules via coherent optical interfaces or physical exchange of qubits via matter links.
These mechanisms establish a hierarchical connectivity structure, from in-chain reordering via intra-node shuttling to nodal interconnects, where each added connectivity mechanism entails an additional layer in the compiler stack~\cite{jengModularCompilationQuantum2025}.

As summarized in \Cref{tab:archetypes}, the architectural archetypes mainly differ along a few design axes, namely the number and capacity of processing zones, the choice between global and addressed control, the 1D or 2D register layout, the presence of junctions, and the type of nodal interconnects. The archetypes described below are ordered along that hierarchy: each adds one connectivity mechanism to the one above it. \Cref{tab:archetypes} carries the same archetypes as rows, with (c) and (d) sharing one, so the list below and the table are best read side by side.

\begin{enumerate}[align=left, widest={(c,\,d)}, leftmargin=*]
    \item[(a)] \textbf{Single-well, all-to-all architectures.}
    Static chains are stored in traps with simple electrode structures~\cite{eganFaulttolerantControlErrorcorrected2021, pogorelovCompactIonTrapQuantum2021}, and laser addressing is used to enable all-to-all connectivity.
    The operational fidelities degrade as the register grows.
    The compiler stack needs to find suitable gate decompositions while suppressing crosstalk errors. 
    \item[(b)] \textbf{Multi-well architectures.}
    Small ion chains are stored in adjacent potential wells~\cite{hilderFaultTolerantParityReadout2022}; gate and SPAM operations take place in a fixed processing zone.
    A shuttling layer within the compilation stack schedules shuttling operations to enable effective connectivity~\cite{durandauAutomatedGenerationShuttling2023}. 
    \item[(c,\,d)] \textbf{Multi-well, multi-zone architectures.}
    Gate operations run in parallel in multiple processing zones and shuttling operations establish full connectivity~\cite{mosesRaceTrackTrappedIonQuantum2023}.
    The approach may be combined with addressing to increase the size of fully connected subregisters.
    This approach puts stringent demands on the routing and placement layer.
    Scalability may be increased by adding junctions~\cite{ransford98qubitTrappedionQuantum2026}.
    \item[(e)] \textbf{Multi-zone 2D architectures.} These operate on a medium- to large-scale planar array of interleaved storage and processing zones~\cite{lekitschBlueprintMicrowaveTrapped2017}.
    Arbitrary shuttling control would require a control and wiring overhead on the order of the qubit count, so that multiplexed electrode control has to be employed~\cite{malinowskiHowWire1000Qubit2023, jonesWISERSystematicDesignSpace2026}.
    This requires different approaches to circuit partitioning, routing/placement, and shuttling.
    Fault-tolerant operation requires syndrome extraction across all logical qubits within the coherence time, so that the number of processing zones has to scale with the qubit count~\cite{jonesArchitectingScalableTrapped2026}, see \Cref{sec:fault-tolerance} for what this demands of the compiler and \Cref{sec:teleportation} for what added zones are actually worth on one such routine.
    \item[(f)] \textbf{Modular platforms.} These operate on connected nodes with different types of interconnects, either optical links or matter links~\cite{kimModularUniversalScalable2011, monroeLargescaleModularQuantumcomputer2014}.
    Modular architectures are considered one of the leading routes toward application-scale, fully error-corrected machines. Compared with shuttling within one node, routing via interconnects is more costly in terms of timing and error resources. This establishes a hierarchical connectivity that needs to be reflected by the compiler stack.
\end{enumerate}
Architectural design choices entail trade-offs among connectivity, crosstalk, routing overhead, and the overall complexity of the hardware setup. The more complex the platform, the more the compiler stack decides how much of its capability can actually be used. Co-designing the compiler stack with the hardware architecture can therefore guide design choices and provide resource estimates for the intended workloads~\cite{schoenbergerUsingCompilerFrameworks} already at the design stage.
The remainder of this work follows archetypes (b), (c) and (e), which differ most in how connectivity is established, and leaves the single-well and modular ends of the range aside.

\section{Compilation for trapped ions}\label{sec:compilation}
As discussed in the preceding sections, an architecture is represented by data structures containing the platform's capabilities, constraints, and relevant properties. This data is the foundation of the compiler stack, which turns abstract quantum protocols into running computations. The interface between the abstract representation of an architecture and the compiler is where hardware and software developers meet. This section takes the architecture abstractions as given and asks to which compilation problems they are related. Further below, we include a logical compilation level for fault-tolerant platforms.

\subsection{General concepts}\label{sec:general_concepts}
\begin{figure*}[t]
    \centering
    \includegraphics[width=1\linewidth]{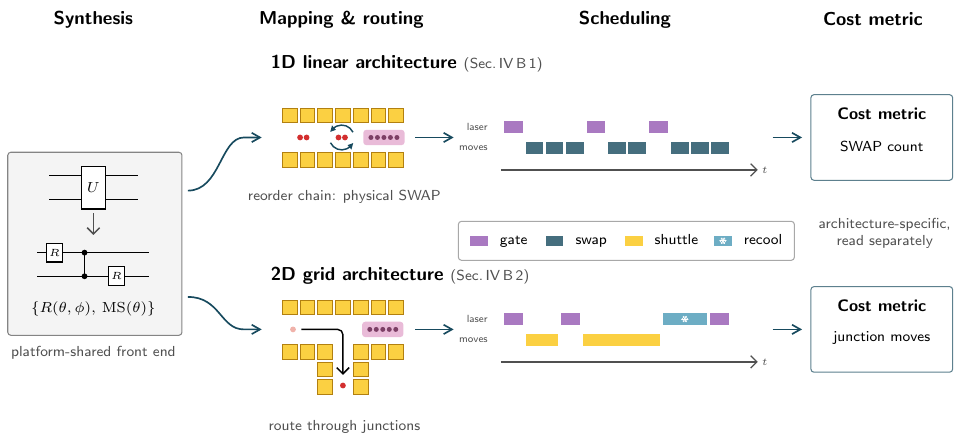}
    \caption{Compilation pipeline for trapped ions: A platform-shared synthesis stage decomposes the circuit into the native gate set (\Cref{sec:operations}), before the architecture-specific mapping and scheduling stages diverge between the two regimes of \Cref{sec:compilation-physical}, summarized in \Cref{tab:compilation-1d-2d}.
    In a 1D architecture (top), connectivity is established, for example, by reordering the ions along the trap axis, and the schedule is dominated by physical SWAPs.
    In a 2D architecture (bottom), ions are routed through junctions, and the schedule is dominated by junction transports. 
    Every SWAP takes the same time, so a reordering appears as a run of equally long bars, whereas a transport takes longer the further the ion is carried.
    The resulting cost metrics are architecture-specific.}
    \label{fig:compilation-steps}
\end{figure*}

Compilation turns an abstract quantum circuit into an executable sequence of hardware operations, organized into successive layers as sketched in \Cref{fig:compilation-steps}:
\begin{description}
    \item[Synthesis] The gates of the circuit are decomposed into the native gate set of \Cref{sec:operations} and the result is optimized to reduce the overall gate count.
    \item[Mapping \& Routing] The circuit qubits are assigned to ions and trap zones, and the operations that bring interacting qubits together are inserted.
    \item[Scheduling] The operations are arranged in time, respecting their dependencies and exploiting the parallelism permitted by the hardware.
\end{description}
For the \emph{synthesis} stage, circuit optimization frameworks developed for other platforms are commonly employed, since these methods do not strongly depend on the qubit implementation or other platform specifics~\cite{kreppelQuantumCircuitCompiler2023, schmaleBackendCompilerPhases2022}. We therefore do not treat synthesis in further detail and concentrate on the platform-dependent stages of mapping, routing, and scheduling.

The trap layout enters at the mapping and scheduling stages, which must respect the connectivity, the number of processing zones, and the junctions abstracted in \Cref{sec:architecture}. In contrast to platforms with fixed connectivity, the mapping stage does not need to insert SWAP gates to establish connectivity. Compilation therefore takes the form of the routing and placement layer of \Cref{sec:shuttling}, which produces a sequence of shuttling operations.

The three compilation stages operate on ions and are part of the physical layer. Quantum error correction adds a logical layer on top, where the objects being synthesized, mapped, and scheduled are code patches rather than ions, and the same three tasks need to be addressed under different constraints (\Cref{sec:fault-tolerance}).

Structuring a stack of this kind is a problem classical compiler construction has already solved once.
MLIR lets each layer be expressed in its own dialect and lowered into the next, which is the shape of the physical and logical layers above. It has been adopted for quantum compilation on that basis~\cite{lattnerMLIRScalingCompiler2021, mccaskeyMLIRDialectQuantum2021, hopfIntegratingQuantumSoftware2026}, though the dialects and lowering passes remain platform-specific. Trapped-ion-specific MLIR dialects are still under development.

\begin{summary}[The compilation problem]
    \textbf{Input:} a circuit as a gate dependency graph, and the architecture data of \Cref{fig:HW2architecture}: connectivity, native gate set, operation timings, and fidelities.\\
    \textbf{Output:} a schedule of native gates, shuttling primitives, and measurements.\\
    \textbf{Objective:} the approximate success probability $\mathcal{P}$ of \Cref{sec:errors}; every metric below is a proxy for it.
\end{summary}

\subsection{Physical-level compilation across architectures}\label{sec:compilation-physical}
This section examines how architectural choices shape the physical-level compilation problem.
Ions have to be shuttled, and the constraints are determined by the trap hardware~\cite{contaToolchainShuttlingTrappedion2026}. The computational complexity depends on the architecture. The last column of \Cref{tab:archetypes} shows representative compiler tools for each archetype.
To emphasize the relation between hardware and computational complexity, we consider two representative archetypes that differ fundamentally in how connectivity is established: the 1D linear multi-zone architecture, archetype (c), where routing is realized by reordering ions along a line, and the 2D grid architecture, archetype (e), where ions are routed through junctions on a graph (see \Cref{fig:architectures}). 
Their routing costs are therefore measured differently, by SWAPs in 1D and junction moves in 2D, but both quantify the architecture-imposed routing work that contributes to the success probability through operation fidelities and idle errors (\Cref{tab:compilation-1d-2d} and \Cref{sec:errors}).

\begin{table}[tb]
    \centering
    \caption{Comparison of the hardware models, routing operations, cost metrics, and solution methods used for physical-level compilation in the 1D linear and 2D grid architectures.}
    \label{tab:compilation-1d-2d}
    \renewcommand\cellalign{tl}
    \renewcommand\theadalign{tl}
    \setlength{\tabcolsep}{5pt}
    \renewcommand{\arraystretch}{1.2}
    \footnotesize
    \begin{tabular}{@{}lll@{}}
        \toprule
        & \thead{1D linear\\architecture} & \thead{2D grid\\architecture} \\
        \midrule
        \makecell{Hardware\\model} & \makecell{linear order of\\ion positions} & \makecell{junction/position\\graph over zones} \\
        \addlinespace
        \makecell{Routing\\move} & \makecell{physical SWAP,\\split/merge, transport\\along the axis} & \makecell{move, split/merge,\\junction transport\\between zones} \\
        \addlinespace
        \makecell{Cost\\metric} & SWAP count & junction moves \\
        \addlinespace
        \makecell{Solution\\methods} & \makecell{gate aggregation\\$+$ greedy/tabu\\search} & \makecell{graph scheduling\\or exact SAT} \\
        \bottomrule
    \end{tabular}
\end{table}

\subsubsection{1D linear architecture: reordering and SWAP-based compilation}
\textbf{Setting.} This is archetype (c) of \Cref{fig:architectures}, drawn as the upper lane of \Cref{fig:compilation-steps}.
In a 1D linear architecture, the ions sit in a fixed order along the trap axis.
Splitting, merging, and transport can bring a pair into a processing zone together, but reaching an arbitrary pair means reordering ions.
A two-qubit gate can be applied once its operands have been brought together in a processing zone, so every gate whose qubits are not already adjacent requires a reordering of the ion positions consisting of possibly multiple operations.

\textbf{Pipeline.} Once synthesis is complete (\Cref{sec:general_concepts}), mapping and scheduling reduce to finding a sequence of reorderings that makes the operands of each two-qubit gate adjacent while minimizing the required exchanges.
A reordering is performed by swapping neighboring ions, each realized by rotating a two-ion crystal. 
The minimum number of such swaps that realize a permutation $\pi$ of the register equals its number of inversions,
\begin{equation}\label{eq:swap-inv}
    N_{\mathrm{SWAP}}(\pi) = \mathrm{inv}(\pi),
\end{equation}
the count of qubit pairs whose order $\pi$ reverses.
The routing cost is thus set by how far the required permutations displace the qubits.
The compiler evaluated in \Cref{sec:case_studies} groups the gates into successive processing steps (\Cref{fig:mainzcompiler_exp}) and minimizes a distance-based surrogate of this cost; its cost function and search strategy are deferred to that section.

\textbf{Compilers.} The tools listed for archetypes (b) and (c,\,d) in \Cref{tab:archetypes} differ mainly in how they search for this ordering, and most of them search heuristically.
Kreppel et al.~\cite{kreppelQuantumCircuitCompiler2023} build on the \texttt{pytket} framework~\cite{sivarajahTketRetargetableCompiler2021} and remove the SWAP gates before mapping, reintroducing them as physical ion exchanges, while Durandau et al.~\cite{durandauAutomatedGenerationShuttling2023} generate the shuttling sequences for a linear segmented trap directly and compare several strategies for the initial qubit assignment.
Learning-based alternatives fine-tune a language model to find the shuttling operations, for linear as well as branched layouts~\cite{kreppelShuttlingCompilerTrappedIon2025}, or learn a shuttling policy by reinforcement learning~\cite{schierReinforcementLearningIon2026}.

\textbf{Metric.} 
The reported cost is the SWAP count, which reflects routing overhead most directly for a filled register, where each SWAP reorders data-carrying ions instead of moving idle ions aside. Split, merge, and transport operations are not counted, so the metric is an architecture-specific proxy and not a complete execution cost.

\subsubsection{2D grid architecture: shuttling and junction routing}\label{sec:compilation_for_qccd}
\textbf{Setting.} This is archetype (e) of \Cref{fig:architectures}, drawn as the lower lane of \Cref{fig:compilation-steps}.
In a two-dimensional grid, the zones are connected by junctions, so an ion can be routed through a junction to pass another instead of waiting for it, which a single line of wells cannot do.
The compiler, therefore, no longer sorts ions along a line but routes them across a graph, and the junctions that cross between segments are the slow, contended operations it must use sparingly.

\textbf{Pipeline.} Mapping now partitions the circuit across the zones and builds a shuttling schedule that brings each two-qubit gate's operands into a common processing zone.
The routing cost associated with bringing two interacting ions together is determined by their distance in the junction graph $G$,
\begin{equation}
    c(i \to j) = d_G(v_i, v_j),
\end{equation}
the number of segment and junction hops separating their positions $v_i$ and $v_j$; where the one-dimensional cost was quantified by the inversion count (\Cref{eq:swap-inv}), the two-dimensional cost is a distance on a graph.
The cost counts the junction transports needed to bring the two ions together, and how many of them can run in parallel is set by the junctions the competing routes share.
The compiler reduces this cost by routing ions along shortest paths, using the cycles of the graph to avoid backtracking, and by choosing the next gate to execute at runtime~\cite{schoenbergerShuttlingScalableTrappedIon2024}.

\textbf{Compilers.} For archetype (e) the same problem has been addressed both heuristically and exactly.
Graph-based heuristic scheduling drives the \texttt{mqt-ion-shuttler}~\cite{schoenbergerShuttlingScalableTrappedIon2024, schoenbergerOrchestratingMultiZoneShuttling2025}; for provably minimal schedules, the routing problem can be formulated as a Boolean satisfiability problem and solved exactly, although this approach is limited to small instances~\cite{schoenbergerUsingBooleanSatisfiability2024}; and the position-graph abstraction ports the mature qubit-mapping heuristics of superconducting compilers to the shuttling problem~\cite{bachEfficientCompilationShuttling2026}.
More scalable approaches include the SIMD-aware TrapSIMD~\cite{ruanTrapSIMDSIMDAwareCompiler2025a}, the multi-level scheduler MUSS-TI~\cite{wuMUSSTIMultilevelShuttle2025}, and a parallelism-aware router that balances zone occupancy against ion movement~\cite{ovideAdaptiveParallelismAwareQubit2026}.

\textbf{Metric.} Two counts are in use here.
The number of \emph{junction moves} totals the individual ion transports through a junction, while the number of \emph{shuttling time steps} gives the length of the resulting schedule, within which moves that share neither electrodes nor a junction proceed together.
Depending on operation times and fidelities, either count can be a better proxy for the routing problem and cost functions like the error model of \Cref{sec:errors} can be used to combine them.

\begin{summary}[Physical-level compilation]
    \textbf{Task:} place the circuit qubits on ions and zones and order the moves that bring each gate's operands into one processing zone.\\
    \textbf{Constraint:} two qubits interact only within one crystal, and moves sharing electrodes or a junction serialize.\\
    \textbf{Metric:} elementary operation counts, execution time or combinations thereof, as a proxy for the success probability of \Cref{sec:errors}.
\end{summary}

\subsection{Compilation for fault-tolerant quantum computing}\label{sec:fault-tolerance}
To connect physical-level compilation to fault-tolerant operation, we now consider the additional compilation constraints introduced by quantum error correction.
Computation at scale requires the machinery of quantum fault tolerance~\cite{campbellRoadsFaulttolerantUniversal2017, bermudezAssessingProgressTrappedIon2017}.

Trapped-ion systems have contributed to quantum error correction through repeated stabilizer readout and fault-tolerant universal gate sets~\cite{hilderFaultTolerantParityReadout2022, postlerDemonstrationFaulttolerantUniversal2022}, mid-circuit capabilities and real-time feedforward~\cite{ryan-andersonRealizationRealTimeFaultTolerant2021}, single-shot error correction and fault-tolerant code switching~\cite{postlerDemonstrationFaultTolerantSteane2024, pogorelovExperimentalFaulttolerantCode2025, daguerreExperimentalDemonstrationHighFidelity2025}, and, more recently, beneficial quantum error correction beyond break-even~\cite{hongEntanglingFourLogical2024, paetznickImprovedQuantumProcessor2026, berthusenExperimentalValidationCompact2026}.

These demonstrations operate at the scale of a few logical qubits, where encoded circuits can still be laid out by hand. Beyond this scale, the encoded level becomes an additional compilation problem in which architecture and compiler must again be considered jointly. As at the physical level, a logical circuit must be synthesized, mapped, and scheduled, but the gate set, the objects being placed, and the cost model change.

For illustration, we use a single logical qubit per code patch and the common Clifford+$T$ universal gate set as a running example. More advanced schemes, such as the recent cat-state-based approach~\cite{tripierFaultTolerantQuantumComputing2026} and fast surgery operations for qLDPC codes~\cite{baspinFastSurgeryQuantum2025}, illustrate the broader range of fault-tolerant compilation strategies.

Fault-tolerant compilation constrains the stack at two levels. Physically, syndrome extraction runs throughout the computation and needs mid-circuit measurement and real-time classical feedback. At the logical level the compiler works on encoded qubits and a fault-tolerant gate set, which brings new synthesis, mapping, and scheduling objectives.

\subsubsection{Syndrome extraction and mid-circuit measurements}\label{sec:fault-tolerance-compilation}
\begin{figure}[tb]
    \centering
    \includegraphics[width=\columnwidth]{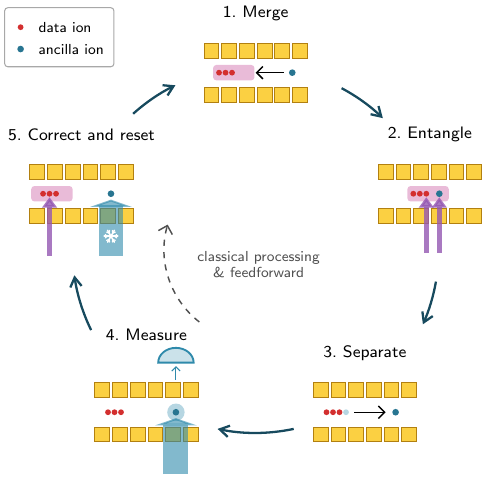}
    \caption{
    One round of syndrome extraction on a trapped-ion register, drawn in the operation vocabulary of \Cref{fig:capabilities}.
    The ancilla ion (blue) is merged with the data ions (red) into a processing zone and entangled with them through the native gate set.
    It is then separated from them again by shuttling, because fluorescence readout is destructive and global on a crystal (\Cref{sec:measurement}), and measured mid-circuit.
    The outcome is decoded classically and fed forward in real time. This is the inner arc, which runs from the measurement to the step it conditions: an addressed gate applies the correction to the data ions, while the ancilla is reset and recooled where it was measured.}
    \label{fig:qec-round}
\end{figure}

At the physical level, quantum error correction repeatedly invokes the routing and scheduling problem of \Cref{sec:compilation-physical}: the computation becomes a long schedule of syndrome-extraction rounds, each interleaved with mid-circuit measurement and real-time feedback.
\Cref{fig:qec-round} draws one such round as a cycle: the outer ring carries the quantum operations listed below, while the inner arc runs counter to it and carries the classical feedback, which has to close before the next round can begin.
\begin{description}
    \item[Ancilla preparation] the ancilla ions are reset to a fixed state, an instance of the state preparation of \Cref{sec:measurement}.
    \item[Stabilizer entangling] each ancilla is entangled with the data qubits of its stabilizer through native two-qubit gates, which first requires bringing them together, the routing problem of \Cref{sec:compilation-physical}, now repeated every cycle.
    \item[Mid-circuit readout] the ancillas are measured without disturbing the data.
    Since readout is destructive and, on a shared crystal, disruptive to the neighboring ions, the measured ion must be isolated, either shuttled to a dedicated readout zone or read out through metastable or second-species states.
    \item[Decode and correct] the syndrome is passed to a decoder and the correction is applied, or tracked in software, in real time and within the coherence budget.
\end{description}

Unlike an arbitrary circuit, a syndrome-extraction round repeats a known shape each cycle: a bipartite data-ancilla interaction graph and two-qubit gates of a single kind.
Compilers written for this task exploit that regularity, for example by shuttling only the ancilla ions or only the data ions to reduce movement~\cite{khanMovelessMinimizingQEC2025} and by reusing ion positions across rounds through mid-circuit measurement and reset~\cite{decrossQubitReuseCompilationMidCircuit2023}.

\textbf{Architectural constraints.} How well the round can be scheduled depends on the architecture.
The number of processing zones and the way they are connected fix how many stabilizers can be measured at once. The trap layout is therefore part of the code design~\cite{khanCycloneDesigningEfficient2026}, and the encoding can be adapted to the workload in situ~\cite{yinISwitchQECDemand2026}.
The surface-code compilers of \Cref{tab:archetypes} translate a given code and device into the concrete shuttling schedule for one round~\cite{leblondTISCCSurfaceCode2023, jonesArchitectingScalableTrapped2026}.

\subsubsection{Logical level compilation}\label{sec:logical-routing}

At the logical level, the same three compilation stages reappear under different primitives: a logical circuit is synthesized into a fault-tolerant gate set, encoded qubits are mapped and routed across the device, and logical operations are scheduled.

\textbf{Logical entangling gates.} 
CSS codes admit a transversal CNOT, allowing logical entangling gates to be implemented by acting on each physical qubit individually or, alternatively, through lattice surgery~\cite{litinskiGameSurfaceCodes2019, ryan-andersonHighfidelityTeleportationLogical2024}. In two-dimensional nearest-neighbor architectures, one usually has to rely on patch-based routing and lattice surgery. QCCD architectures provide an alternative: in-zone connectivity and ion shuttling can bring the two code blocks into a common interaction region, enabling the efficient transversal implementation.

\textbf{Non-Clifford gates.}
No quantum error-correcting code admits a universal transversal gate set~\cite{eastinRestrictionsTransversalEncoded2009}, so universal fault-tolerant computation requires additional mechanisms. The first fault-tolerant universal gate set demonstrated with trapped ions used magic-state injection by teleportation~\cite{postlerDemonstrationFaulttolerantUniversal2022}. Several recent demonstrations switch codes instead, which avoids a separate distillation factory at the small code distances considered and reuses native operations~\cite{pogorelovExperimentalFaulttolerantCode2025, daguerreExperimentalDemonstrationHighFidelity2025, buttFaultTolerantCodeSwitchingProtocols2024}.

Compilers for the logical layer already exist. They lay out the code patches, route them e.g.\ via lattice surgery, and schedule the consumption of magic states, at circuit sizes reaching millions of gates~\cite{watkinsHighPerformanceCompiler2024}.
Fault-tolerant execution of complete algorithms has recently been demonstrated on ion hardware~\cite{perlinFaulttolerantExecutionErrorcorrected2026, nelsonBelowthresholdExecutionFaulttolerant2026}.

\begin{summary}[Fault-tolerant compilation]
    \textbf{Task:} schedule repeated syndrome extraction at the physical level and synthesize, place, route, and schedule encoded operations at the logical level.\\
    \textbf{Constraint:} ancillas must be isolated for readout and subsequently recooled, while classical decoding must complete within the coherence budget.\\
    \textbf{Metric:} physical syndrome-round duration and logical resource costs, such as non-Clifford resource consumption or routing overhead, depending on the compilation layer.
\end{summary}

\section{Case studies and discussion}\label{sec:case_studies}
We demonstrate the interplay between architecture and compiler by compiling two representative workloads for the linear archetypes (b) and (c) and the two-dimensional archetype (e) of \Cref{sec:compilation-physical} and comparing the resulting architecture-dependent routing overheads. The first uses random circuits to study how routing cost scales with register size, a central consideration when comparing hardware designs~\cite{schoenbergerUsingCompilerFrameworks}. The second uses a logical teleportation circuit as a structured fault-tolerant workload to study how the same architectural choices affect routing overhead in a concrete application.

\subsection{Compiler implementations}\label{sec:case-compilers}
The case studies below run the same circuits through two compilers, one for each of the regimes of \Cref{sec:compilation-physical}.
Both follow the stages of \Cref{sec:general_concepts} and differ in how they realize mapping and scheduling.

\begin{figure}[tb]
    \centering
    \includegraphics[width=\linewidth]{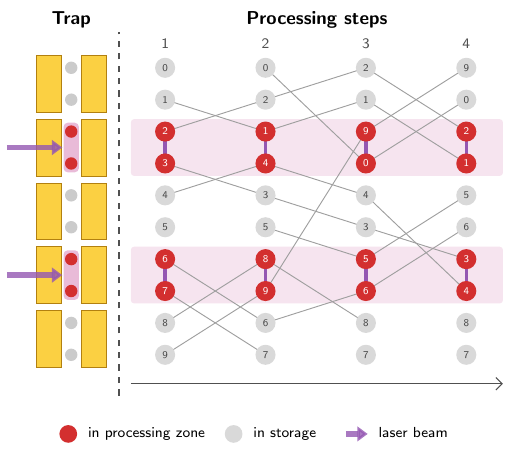}
    \caption{
    Aggregated arrangement produced by the one-dimensional compiler for a linear multi-zone architecture (type (c) in \Cref{fig:architectures}).
    The trap on the left fixes the positions a qubit can occupy, and two of its segments are processing zones, each holding the pair of ions a gate acts on.
    Each column on the right is one processing step.
    Aggregation chooses which qubits occupy those zones at each step, shown in red, which is what makes the indicated two-qubit gates executable in parallel; the remaining qubits wait in storage, shown in grey.
    Between consecutive steps the connecting lines mark the ions that change position, and their travel is the quantity minimized by \Cref{eq:cost_1d}.}
    \label{fig:mainzcompiler_exp}
\end{figure}

\textbf{Linear-architecture compiler.} For the linear segmented trap we use a routing algorithm developed for this work.
Executing a circuit on this architecture requires mapping it onto the physical layout of the trap, which has two closely coupled aspects: an assignment of qubits to the architecture and the movement of ions within it.
They are treated in turn:
\begin{description}
    \item[Aggregation] sets of qubits are identified that maximize the number of gates executable in one processing step.
    The result is a sequence of arrangements of the kind shown in \Cref{fig:mainzcompiler_exp}.
    \item[Movement optimization] the arrangement is then optimized with respect to the total ion displacement over the circuit, thereby reducing the SWAP count.
\end{description}
The second stage minimizes the cost $\Lambda$, for a circuit $\mathcal{C}$ and an arrangement $\mathcal{A}$ of qubits in aggregated processing blocks,
\begin{equation}\label{eq:cost_1d}
    \Lambda_{\mathcal{C}}(\mathcal{A})=\sum_{d=1}^{D_L-1} \Lambda(S^{(d)},S^{(d+1)}),
\end{equation}
where the sum runs over successive pairs among the $D_L$ processing steps $S^{(d)}$ of the arrangement.
In \Cref{fig:mainzcompiler_exp} these steps are the columns, and the quantity being summed is the total length of the connecting lines between two neighboring columns.
The rearrangement cost between two such steps is
\begin{equation}
    \Lambda(S^{(d)},S^{(d+1)})=\sum_q \bigl\vert X_q^{(d)}-X_q^{(d+1)}\bigr\vert^{\kappa},
\end{equation}
with $X_q^{(d)}$ the position of qubit $q$ along the trap axis in step $d$ and $\kappa$ an exponent controlling the relative penalty assigned to long displacements.
The minimization combines a greedy algorithm with tabu search, including stochastic search steps. The final SWAP count is obtained by summing the inversion counts between consecutive optimized qubit arrangements.

The cost model includes only SWAP overhead, while split, merge, and transport costs are deliberately neglected. Consequently, only two-qubit gates enter the optimization; single-qubit gates are assumed to be scheduled adjacent to a two-qubit gate on the same qubit, with any additional split/merge overhead excluded from the cost model.

\textbf{2D-compiler.} For the grid we use the multi-zone shuttling compiler of the Munich Quantum Toolkit~\cite{schoenbergerOrchestratingMultiZoneShuttling2025}, which places the processing zones outside a grid-shaped memory region and models their internal structure instead of treating them as black boxes.
It partitions the circuit qubits across the zones and selects the gates to execute next from the gate dependency graph, so that several zones can be kept busy at the same time, while the ions themselves are routed along the shortest paths and cycles of the junction graph of \Cref{sec:compilation_for_qccd}~\cite{schoenbergerShuttlingScalableTrappedIon2024}.
It reports both counts defined in \Cref{sec:compilation_for_qccd}: shuttling time steps and individual junction moves.
We report the number of junction moves below, since the number of transport operations provides the closest counterpart to the SWAP count of the linear architecture.

\subsection{Evaluated architectures}\label{sec:case-architectures}
The compilers of \Cref{sec:case-compilers} are paired with a small set of architectures, chosen so that the design axes of \Cref{sec:architecture} are varied one at a time rather than all at once.
We characterize each configuration by its architectural parameters:
\begin{description}
    \item[Processing zones] one or two for the linear architecture, one to four for the grid.
   
    \item[Zone capacity and addressing] two ions without addressing, so that a zone holds exactly the operands of one gate, or ten ions with individual addressing, so that a zone holds a small sub-register and any pair inside it can be selected (\Cref{sec:2Q_gates}).
    The two properties are varied together because there is no benefit to addressing in a two-ion zone.
\end{description}
The linear configurations correspond to the multi-well and multi-zone archetypes of \Cref{tab:archetypes}, while the grid corresponds to the multi-zone 2D archetype. All evaluated configurations remain within the register sizes demonstrated by current devices.

Both case studies use this configuration set, with the varied design parameters specified separately below.

\subsection{Random circuits}
\label{sec:random-circuits}
The two compilers count different quantities, so this first case study focuses on scaling rather than on quantitative comparison.

\textbf{Circuits.} Each random circuit consists of $L$ sequential interaction layers.
A layer is generated by sampling a random permutation of the qubit indices and pairing adjacent elements, resulting in $\lfloor N/2 \rfloor$ disjoint two-qubit interactions that can be executed in parallel.
The total number of two-qubit gates is therefore
\[
m(N) = L \,\lfloor N/2 \rfloor.
\]
To maintain comparable global connectivity as the system size increases, the number of layers is chosen to grow logarithmically with the number of qubits, $L \propto \log N$.

\textbf{Architectures.} All four linear configurations of \Cref{sec:case-architectures} are evaluated.
On the grid, we keep two processing zones and increase the grid size along with the register, so that the device is never much larger than the ion number requires.
The linear architecture has fixed zone capacities and is extended incrementally as qubits are added. With two processing zones, the processing zones fill first, then the storage zones between them, then the outermost ones. With a single processing zone, the storage zones fill symmetrically around it.

\begin{figure*}[t]
    \centering
    \includegraphics[width=\linewidth]{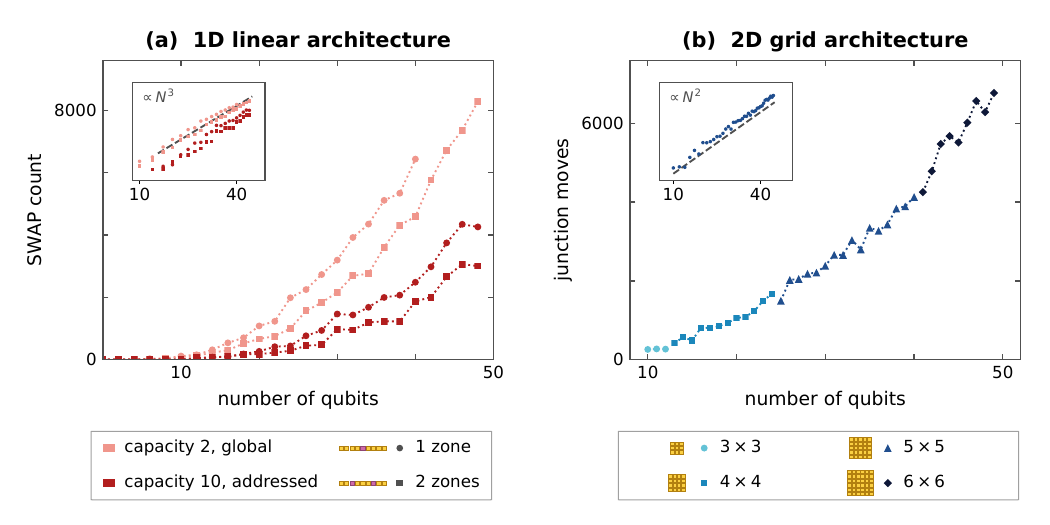}
    \caption{Scaling of the routing cost with register size.
    (a) Four configurations of the linear architecture, identified by the number of processing zones, the zone capacity, and whether the ions are addressed individually. The legend depicts a miniature of each trap.
    (b) The two-dimensional grid, which is enlarged as the register grows, with a miniature of the memory region next to the segment it belongs to.
    The insets show the same data over the same axes, replotted on a double-logarithmic scale together with a reference power law.}
    \label{fig:scaling-random}
\end{figure*}

\textbf{Results.} \Cref{fig:scaling-random} shows the two families side by side, each panel with its own metric.
In both architectures, the routing cost increases with register size: larger circuits contain more interactions, while the interacting qubits must on average be routed over greater distances.

Within the linear architecture (\Cref{fig:scaling-random} (a)), the two design axes act differently.
A second processing zone lowers the SWAP count at almost every size by allowing the compiler to distribute gates across both zones and reduce ion rearrangement.
Enlarging the zones to ten ions and addressing them individually lowers it considerably further, and is the more effective of the two in this case: a gate no longer requires its operands to be alone in the zone, thereby substantially reducing the required ion reordering. But larger addressed zones are not free in practice, since crosstalk is affected by the number of simultaneously addressed ions (\Cref{sec:2Q_gates}) and the choice of zone capacity carries its own architectural trade-offs~\cite{muraliArchitectingNoisyIntermediateScale2020, jonesArchitectingScalableTrapped2026}.
Still, this demonstrates an interesting trade-off between two hardware design choices.

On the grid (\Cref{fig:scaling-random} (b)) the cost also grows with the register.
The visible steps are due to the memory region being enlarged in four stages as the register fills, from $3\times3$ to $6\times6$ to always fit the corresponding number of qubits.

Over the register sizes studied, the log-scale insets indicate approximately quadratic growth for the grid and approximately cubic growth for the linear architecture. Increasing zone capacity and addressability shifts the curves without markedly changing their slope. This difference is largely geometric. For a roughly uniform $d$-dimensional layout, $N$ ions occupy a region with linear extent proportional to $N^{1/d}$, so typical routing distances grow more rapidly in a line than on a grid. In addition, a 1D layout provides no alternative path when two ions must pass one another, requiring SWAP operations that also displace other ions, whereas a 2D grid provides additional routing paths through junctions.

\subsection{Logical teleportation circuit}\label{sec:teleportation}
\begin{figure*}[t]
    \centering
    \includegraphics[width=\linewidth]{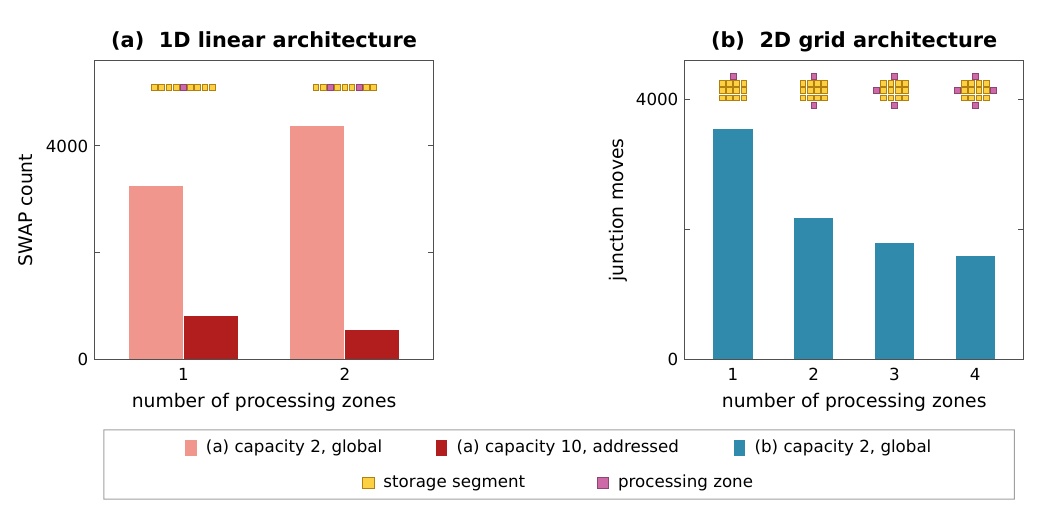}
    \caption{Routing cost for the $41$-qubit logical teleportation circuit.
    The configurations of \Cref{sec:case-architectures} are identified by their properties: the number of processing zones on the horizontal axis, zone capacity and addressing by color, and the layout by the panel, (a) the 1D linear architecture and (b) the 2D grid architecture, which runs at capacity 2 without addressing throughout.
    Above each bar, a miniature shows the corresponding topology in the vocabulary of \Cref{fig:architectures}, a row of storage segments for the 1D linear architecture and a grid memory region with the processing zones attached outside it for the 2D grid architecture.}
    \label{fig:teleportation}
\end{figure*}

The second case study uses a structured logical teleportation circuit as a fault-tolerant workload.
It is also the logical primitive of \Cref{sec:logical-routing}, executed here in its transversal form.

We encode each logical qubit using the seven-qubit Steane code~\cite{steaneSimpleQuantumErrorcorrecting1996}. The three logical qubits therefore form three seven-ion blocks, and each logical two-qubit gate becomes a transversal layer acting across two blocks.
The logical CNOT gates are implemented transversally on the physical qubits, while the logical Bell-basis measurements accumulate the parity of each block onto an ancilla.
The subsequent Pauli corrections ($X$ and $Z$) are conditioned on the Bell-measurement outcome.
The resulting circuit uses 41 qubits and 100 two-qubit gates. 21 of the qubits make up the three logical qubits and the remainder are ancillas for syndrome extraction and logical parity checks.

\textbf{Architectures.} 
The circuit has fixed width, so here we vary the number of processing zones instead of the register size: one or two for the linear architecture at both zone capacities, and one to four for the grid at a fixed memory-region size. This comparison tests whether routing overhead is reduced more effectively by adding processing zones or by increasing their capacity and addressability.

\textbf{Results.} \Cref{fig:teleportation} collects the routing cost across the configurations.
The same two axes behave differently on this workload.
Enlarging the zones to ten ions and addressing them individually reduces the SWAP count by a factor of four with one processing zone and by nearly eight with two.
Adding a second processing zone, in contrast, only helps once the zones are large: at capacity ten it saves a further third, whereas at capacity two it \emph{increases} the SWAP count by about a third. Fixing the number of qubits held between the two zones removes routing flexibility, and at capacity two that loss appears to outweigh the benefit of the additional zone. 
So this might partially reflect a limitation of the routing heuristic rather than an intrinsic disadvantage of additional processing zones, but a similar non-monotonicity has been reported independently for racetrack-shaped processors~\cite{jangScalableGateLevelParallelism2026}.
Together, these illustrate the importance of co-designing hardware choices and compilation methods. 

On the 2D grid, panel (b), the cost falls monotonically as processing zones are added, with diminishing returns: the second zone removes more than a third of the cost, the third and fourth only a further sixth and tenth.
Also, here, part of this reduction may stem from the compiler's static partitioning rather than from the architecture.
But still, depending on manufacturing complexity, this provides useful data for the trade-off regarding how many processing zones to add to the chip.

\subsection{Discussion}\label{sec:case-discussion}

The case studies do not rank the architectures against one another. They show how architecture and compiler strategy together set the routing overhead, and the two cannot be separated here: each architecture is compiled by one tool, so part of every difference may come from the compiler and not from the architecture. Within the evaluated configurations, two effects dominate. First, in the linear architecture, increasing zone capacity and addressability reduced routing overhead more than adding processing zones did. Second, for the compilers studied, the 2D grid scaled more favorably with register size than the 1D linear architecture.

\section{Conclusion}

This work connects two communities: physics- and hardware-focused researchers investigating how architectural design choices affect device performance, and the computer-science and compiler community developing the software required to exploit the hardware's capabilities. \Cref{sec:platform} collects the platform's native operations with their timing and fidelity costs, in a form that presupposes no atomic physics. \Cref{sec:architecture} organizes the architectures built from them into archetypes that differ in the number and capacity of their processing zones, their degree of addressability, and the presence of junctions. \Cref{sec:compilation} sets out the compilation problem each archetype poses, at the physical and at the logical level. \Cref{sec:case_studies} compiles two workloads for three of the archetypes.

The resulting progression from physical primitives to architecture, compilation, and practical evaluation illustrates why trapped-ion hardware and compiler design cannot be considered independently. In \Cref{sec:case_studies}, the hardware configuration (layout, addressability, zone capacity, and number of processing zones) affected the resulting routing overhead. Questions such as how many processing zones to provide, where to place junctions, and how much addressability to support, therefore, require the joint evaluation of the architecture \emph{and} the compiler. The abstractions and the case-study methodology developed here provide a basis for such hardware-software co-design of future trapped-ion quantum processors.

\begin{acknowledgments}
    This work was funded by the Deutsche Forschungsgemeinschaft (DFG, German
    Research Foundation) -- 563436708, by the BMFTR under grant number
    13N17298 (SYNQ), and by the European Union's Horizon Europe research and
    innovation program under grant agreement No.\ 101114305
    (``MILLENION-SGA1'' EU Project).
    L.S., D.S. and R.W. further acknowledge funding from the European Research
    Council (ERC) under the European Union's Horizon 2020 research and
    innovation program (grant agreement No.\ 101001318), and from the Munich
    Quantum Valley, which is supported by the Bavarian state government with
    funds from the Hightech Agenda Bayern Plus. J.E., J.H., U.P., and F.S.-K. further acknowledge funding by BMFTR within the projects ATIQ and IQuAn. 
    Ch.D.M., J.H., U.P., and F.S.-K. are affiliated with neQxt GmbH, and
    R.W. with MQSC.
    The authors declare no further competing interests.
    During the preparation of this work, the authors used Claude (Anthropic,
    Opus 5) to refine figure designs starting from manual drafts, to prepare the Zenodo archive, and to revise the wording and grammar of the text.
    The authors reviewed and edited the output and take full responsibility for the content of this article.
\end{acknowledgments}

\section*{Data availability statement}
The data that support the findings of this study, together with the scripts
that produce the plotted figures, are openly available on
Zenodo~\cite{schmidEvaluationData2026}.

\bibliography{header,references_short}%

\end{document}